\documentclass[letterpaper]{JHEP3}

\newcommand{\beq}{\begin{equation}}
\newcommand{\eeq}{\end{equation}}
\newcommand{\cA}{{\cal A}}
\newcommand{\cAb}{{\overline{\cal A}}}
\newcommand{\cF}{{\cal F}}
\newcommand{\cFb}{{\overline{\cal F}}}
\newcommand{\cD}{{\cal D}}
\newcommand{\cR}{{\cal R}}
\newcommand{\cDb}{{\overline{\cal D}}}
\newcommand{\cQ}{{\cal Q}}
\newcommand{\cU}{{\cal U}}
\newcommand{\cUb}{{\overline{\cal U}}} 
\newcommand{\KD}{{K\"{a}hler-Dirac }}
\newcommand{\cN}{{\cal N}}
\newcommand{\Tr}{{\rm Tr\;}}
\newcommand{\etab}{{\overline{\eta}}}
\newcommand{\psib}{{\overline{\psi}}}
\newcommand{\phib}{{\overline{\phi}}}
\newcommand{\bx}{{\bf x}}
\newcommand{\bmu}{{\boldsymbol \mu}}
\newcommand{\bnu}{{\boldsymbol \nu}}

\newcommand{\brho}{{\boldsymbol \rho}}

\usepackage{graphicx}
\usepackage{amsmath}
\usepackage{amsfonts}
\usepackage{epsfig}

\title{Anti-de Sitter space from supersymmetric gauge theory}

\author{Simon Catterall \\
Department of Physics, Syracuse University, Syracuse, NY13244, USA \\
E-mail: \email{smc@phy.syr.edu}
}

\preprint{}

\date{December 2008}

\abstract{
We construct a four dimensional Yang-Mills theory 
with ${\cal N}=4$ twisted supersymmetry whose classical
vacua correspond to four dimensional anti-de Sitter space. 
The theory utilizes a complex gauge field whose real
part is a spin connection which is used to
enforce local Lorentz invariance. The imaginary part of
the connection can then be interpreted as a vierbein.
The topological construction 
ensures that the partition function and classical vacua are
independent of the
background geometry. Additionally a supersymmetric and gauge
invariant lattice
construction is possible yielding a non-perturbative definition
of the theory.
}

\begin{document}

%
\section{Introduction}

The problem of constructing a theory of gravity which is consistent
with quantum mechanics has a long history. The non-renormalizability
of quantum general relativity can be traced to the fact that
the Newton constant carries the dimensions of length squared
corresponding to an action linear in the curvature. Actions containing
higher powers of the curvature are usually thought to lead to
violations of unitarity and can at best be thought of
only as effective field theories \cite{Donoghue:1994dn}.

Alternative approaches to gravity have tried to highlight the
similarities to gauge theory, where in the case of gravity, the local
symmetry corresponds to Lorentz invariance. In this case the
corresponding 
gauge field is called a spin connection. It is necessary
to introduce such an object when discussing fermions
in curved spacetime where it is partnered by a new field the
so-called vierbein which describes the local Lorentz frame and
from which the usual metric tensor can be reconstructed.

Einstein's theory can
be reproduced from the spin connection and vierbein
provided a certain additional condition is satisfied; the
vanishing of the torsion, which determines the
spin connection in terms of the vierbein and its derivatives. 
The renormalizability
problems of quantum gravity reemerge in this
approach after this last condition is imposed;
since then the usual Yang-Mills action for the spin connection
becomes fourth order in derivatives - see eg. 
\cite{vanNieuwenhuizen:2004rh}.

A further obstruction to constructing a quantum theory of gravity
is the requirement of background independence -- the result
of quantizing 
the classical theory should not depend on the choice of background
metric. Thus the theory should be topological in character.
 
In this paper we will write down a locally Lorentz invariant
theory in which the spin connection and vierbein are treated
as {\it independent
fundamental fields}. They appear as the real
and imaginary parts of a {\it complex} connection. 
The action contains a conventional
Yang-Mills term which is quadratic in the curvature and
the theory is thus renormalizable. Furthermore the complexified Yang-Mills
Bianchi identity automatically
yields the two gravitational Bianchi identities
of Riemann-Cartan gravity.
In the classical limit the torsion vanishes
and the vierbein and spin connection conspire to yield  
anti-de Sitter space. The construction corresponds to a modification of
a well known twist of $\cN=4$ super Yang-Mills due to Marcus and 
inherits that theory's topological character. Specifically, 
the theory possesses a $\cQ$-exact energy-momentum
tensor where the scalar nilpotent supercharge $\cQ$ appears as
a consequence of the twisting procedure.
This structure ensures that the expectation
value of any $\cQ$-invariant observable is independent of any
background metric used to construct the Yang-Mills theory.

We will argue later that this $\cQ$-invariant and background independent
sector of the model corresponds to a topological supergravity
theory.\footnote{Topological
$\cN=2$ supergravity was considered in \cite{Anselmi:1992tj}}
True background independence in a theory of quantum gravity
would require that {\it all} gauge invariant observables
in the theory
possess vacuum expectation values which are independent of
the background metric. The current construction does
not seem to allow this. However, we are
able to write down a lattice theory which preserves both the
scalar supersymmetry and gauge invariance and which may
represent a partial step in this direction. This lattice model
certainly approximates the continuum theory in a flat background
geometry
and serves as a non-perturbative definition of the theory
in such a background.

We derive the theory in a series of steps; in the next section we
review the Marcus twist of $\cN=4$
super Yang-Mills which is the basis of the construction. In
section~\ref{spinstuff}
we modify the gauge field of this
theory to describe a connection whose real part implements local
$SO(4)$ invariance. We show that the imaginary part of the
connection can be
interpreted as a vierbein and the vacua of the theory correspond
to the sphere $S^4$. In section~\ref{lorentz}
we show that implementing
the same procedure for gauge group $SO(1,3)$ yields
a theory whose vacua correspond to anti-de Sitter space. We note
that similar
arguments can be used to write down a five dimensional model
whose classical vacua correspond to ${\rm AdS}_5$. This forms the basis
for a lattice construction of the four dimensional theory which we
describe. The final
section summarizes our findings and points to future work.

\section{Twisted $\cN=4$ gauge theory}
\label{gauge}
We start from the Marcus twist of 
$\cN=4$ super Yang-Mills \cite{Marcus} in which the bosonic fields of
the model comprise a {\it complex} connection $\cA_\mu,\mu=1\ldots 4$ and
two scalar fields $\phi,\phib$ together with a set of
antisymmetric tensor fields $(\eta,\psi_\mu,
\chi_{\mu\nu},\psib_\mu,\etab)$ representing the sixteen fermion
degrees of freedom\footnote{It is also
the twist of ${\cal N}=4$ YM used in the Geometric Langlands program
\cite{Kapustin:2006pk}} \cite{Unsal:2006qp,Catterall:2007kn}

The action takes the form
\beq
S=\frac{1}{g^2}\left[\cQ\int d^4x\;\sqrt{h}\Lambda-
\frac{1}{2}\int d^4x\;\Tr\epsilon^{\rho\lambda\mu\nu}
\chi_{\rho\lambda}\left(
\cDb_{\left[\mu\right.}\psib_{\left.\nu\right]}
+\frac{1}{2}[\phib,\chi_{\mu\nu}]\right)
\right]\label{action}\eeq
where the $\cQ$-exact term is given by
\beq
\Lambda=
\Tr\left(\chi^{\mu\nu}\cF_{\mu\nu}+\psib^\mu\cD_\mu \phi
+\eta \left([ \cDb^\mu,\cD_\mu ]+[\phib,\phi]\right)-
\frac{1}{2}\eta d\right)
\eeq
and indices are raised and lowered by a background metric $h_{\mu\nu}(x)$.
The nilpotent supersymmetry $\cQ$ acts on the fields as follows
\begin{eqnarray}
\cQ\; \cA_\mu&=&\psi_\mu\nonumber\\
\cQ\; \phi&=&\etab\nonumber\\
\cQ\; \psi_\mu&=&0\nonumber\\
\cQ\; \etab&=&0\nonumber\\
\cQ\; \cAb_\mu&=&0\nonumber\\
\cQ\; \phib&=&0\nonumber\\
\cQ\; \chi_{\mu\nu}&=&-\cFb_{\mu\nu}\nonumber\\
\cQ\; \psib_\mu&=&-\cDb_\mu\phib\nonumber\\
\cQ\; \eta&=&d\nonumber\\
\cQ\; d&=&0
\label{twist}
\end{eqnarray}
The second $\cQ$-closed term is supersymmetric on account of the
Bianchi identities
\begin{eqnarray}
\epsilon^{\rho\lambda\mu\nu}\cD_\lambda \cF_{\mu\nu}&=&0\\
\epsilon^{\rho\lambda\mu\nu}\left[\phi,\cF_{\mu\nu}\right]&=&0\\
\epsilon^{\rho\lambda\mu\nu}\cD_\lambda\cD_\mu\phi&=&0
\label{bianchi}
\end{eqnarray}
The complex covariant derivatives appearing in these expressions are defined
by
\begin{eqnarray}
\cD_\mu&=&\partial_\mu+\cA_\mu=\partial_\mu+A_\mu+iB_\mu\nonumber\\
\cDb_\mu&=&\partial_\mu+\cAb_\mu=\partial_\mu+A_\mu-iB_\mu 
\end{eqnarray}
while in the original construction all fields take values in the adjoint
of a $U(N)$ gauge group\footnote{The generators are taken to be {\it
anti-hermitian} matrices satisfying $\Tr (T^aT^b)=-\delta^{ab}$}.
It should be noted that despite the appearance of a complexified
connection and field strength the theory possesses only the usual
$U(N)$ gauge invariance corresponding to the real part of the connection.

Doing the $\cQ$-variation and integrating out the field $d$ yields
\beq
S=\frac{1}{g^2}\int d^4 x\;\sqrt{h}\left(L_1+L_2+L_3+L_4\right)\eeq
where
\begin{eqnarray}
L_1&=&\Tr \left(
-\cFb^{\mu\nu}\cF_{\mu\nu}+\frac{1}{2}[ \cDb^\mu, \cD_\mu]^2)\right)\nonumber\\
L_2&=&\Tr \left(
2(\cD_\mu\phi)^\dagger(\cD^\mu\phi)+\frac{1}{2}[\phi,\phib]^2\right)\nonumber\\
L_3&=&\Tr \left(
-\chi^{\mu\nu}\cD_{\left[\mu\right.}\psi_{\left.\nu\right]}-
\psib^\mu \cD_\mu\etab-\psi^\mu\cDb_\mu\eta -{\chi^*}^{\mu\nu}
\cDb_{\left[\mu\right.}\psib{\left._\nu\right]}\right)\nonumber\\
L_4&=&\Tr \left(
-\psib^\mu\left[\phi,\psi_\mu\right]-\eta\left[\phib,\etab\right]
-\frac{1}{2}{\chi^*}^{\mu\nu}\left[\phib,\chi_{\mu\nu}\right]\right)
\end{eqnarray}
where $\chi^*$ is the Hodge dual of $\chi$, 
$\chi^{\mu\nu}=\frac{1}{2}\epsilon^{\mu\nu\rho\lambda}\chi_{\rho\lambda}$.
The terms appearing in $L_1$ can then be written
\begin{eqnarray}
\cFb^{\mu\nu}\cF_{\mu\nu}&=&(F_{\mu\nu}-[B_\mu,B_\nu])
                            (F^{\mu\nu}-[B^\mu,B^\nu])+
(D_{\left[\mu\right.}B_{\left.\nu\right]})
(D^{\left[\mu\right.}B^{\left.\nu\right]})\nonumber\\
\frac{1}{2}\left[\cDb^\mu,\cD_\mu\right]^2 &=& -2\left(D_\mu B^\mu\right)^2
\end{eqnarray}
where $F_{\mu\nu}$ and $D_\mu$ denote the usual field strength and
covariant derivative depending on the real part of the connection $A_\mu$.
The classical vacua of this theory correspond to solutions of
the equations
\begin{eqnarray}
F_{\mu\nu}-[B_\mu,B_\nu]&=&0\nonumber\\
D_{\left[\mu\right.}B_{\left.\nu\right]}&=&0\nonumber\\
D_\mu B_\mu+[\phi_1,\phi_2]&=&0\nonumber\\
D_\mu\phi_1-[B_\mu,\phi_2]&=&0\nonumber\\
D_\mu\phi_2-[\phi_1,B_\mu]&=&0\nonumber\\
\label{moduli}
\end{eqnarray}
where $\phi=\phi_1+i\phi_2$.
In \cite{Marcus} this moduli space is argued to correspond to
the space of flat {\it complexified} connections modulo gauge
transformations\footnote{This assumes  that the vacuum value of
the scalars is zero which is clearly one solution to these equations}. 
The topological character of the theory
guarantees that any $\cQ$-invariant observable such as the
partition function can be evaluated {\it exactly} by considering only
gaussian fluctuations about such vacuum configurations. Indeed, 
it is easy to see from eqn.~\ref{action}
that the energy momentum tensor of this
theory is $\cQ$-exact rendering the expectation values of such
topological observables independent of smooth deformations
of the background metric $h_{\mu\nu}(x)$.

Returning to the bosonic action and integrating 
by parts we find that the term linear in $F_{\mu\nu}$ cancels
and the contribution of $L_1$ reads
\beq
L_1=\Tr \left(F_{\mu\nu}F^{\mu\nu}+
2B^\mu D^\nu D_\nu B_\mu-[B_\mu,B_\nu][B^\mu,B^\nu]-2R^{\mu\nu}B_\mu
B_\nu\right)
\label{final}\eeq
where $R_{\mu\nu}$ is the background Ricci tensor.
Thus at strong coupling the bosonic theory possesses the usual
Yang-Mills field strength for a real
$U(N)$ gauge field together with
four massless vectors arising from the imaginary part of the connection
and an additional two scalar fields. The fermionic sector comprises a
\KD field arising from twisting the four Majorana fermions of
$\cN=4$ Yang-Mills. In flat space the theory is fully
equivalent to the usual $\cN=4$ theory as this twisting operation can
be regarded merely as an exotic change of variables -- the massless
vectors can combined with the two scalars to yield the usual
six scalars of $\cN=4$ Yang-Mills and the \KD action is
equivalent to the Dirac action for four degenerate
Majorana spinors. In this case
we can lift the restriction to the topological subsector and
consider the set of all gauge invariant operators as spanning the
space of physical observables.

\section{A generalized spin connection}
\label{spinstuff}
The
appearance of a complex connection naturally lead one to try to find a
relationship to gravity \cite{Ashtekar:1986yd}. Since the Yang-Mills
theory contains no symmetric tensor that could
play the role of a metric we will instead focus
on alternative formulations of gravity which utilize
the language of vielbeins and spin connections. As the latter
function as the gauge field for local Lorentz transformations we
are led to examine a generalized 
complex spin connection of the form
\beq
\cA_\mu(x)=\omega^{\ ab}_\mu(x)\sigma^{ab}+\frac{i}{l_G}e^{\ a}_\mu(x)\gamma^a
\label{spin}\eeq
where the indices $a,b=1\ldots 4$ correspond to indices in
a flat Euclidean tangent space and
we employ the definition 
\beq \sigma^{ab}=\frac{1}{4}[\gamma^a,\gamma^b]\eeq
With the Euclidean gamma matrices taken {\it anti-hermitian}
the corresponding tangent space metric is just
\beq\frac{1}{2}\{\gamma^a,\gamma^b\}=\eta^{ab}=-\delta^{ab}\eeq 
The set of matrices $(\sigma^{ab},\gamma^c)$ obey the additional
algebra
\beq
[\sigma^{ab},\gamma^c]=\delta^{ca}\gamma^b-\delta^{cb}\gamma^a
\label{alg}\eeq
It is not hard to show that this set of $10$ matrices can be taken as
a basis for the algebra $so(5)$. However, by construction,
the action will only be locally invariant
under the real part of this connection corresponding
to the group $SO(4)$.
Notice that we have also inserted an explicit length 
scale $l_G$ into the definition so that the function 
$e^a_\mu$ is dimensionless.

To see that $e^{\ a}_\mu$ can be identified as a vierbein let us examine
the variation of $e_\mu=\gamma^a e_\mu^{\ a}$ under an
infinitesimal (adjoint) gauge transformation 
$\delta e_\mu=[\lambda,e_\mu]$. Using the algebra given in
eqn.~\ref{alg}
we find
\beq
\delta e^{\ a}_\mu= \lambda^{ab}e^{\ b}_\mu\eeq
Thus $e^{\ a}_\mu$ transforms correctly as a fundamental field under local
Lorentz transformations.
In a similar way the covariant derivative $D_\mu e_\nu=\partial_\mu
e_\nu+[\omega_\mu,e_\nu]$
leads to the appropriate covariant derivative of the vierbein $e_\mu^a$
\beq
D_\mu e_\nu^{\ a}=\partial_\mu e^{\ a}_\nu+\omega^{\ ab}_\mu e_\nu^{\ b}\eeq
The gauge transformation of the spin connection is the usual
one
\beq
\delta
\omega^{\ ab}_\mu=\partial_\mu\lambda^{ab}+\omega_\mu^{\ ac}\lambda^{cb}+
\omega_\mu^{\ bc}\lambda^{ac}\eeq
while the curvature of the spin connection is given by the
Yang-Mills field strength
\beq
R_{\mu\nu}^{\ \ ab}=\partial_\mu\omega_\nu^{\ ab}-\partial_\nu\omega_\mu^{\ ab}+
\omega_\mu^{\ ac}\omega_\nu^{\ cb}-\omega_\nu^{\ ac}\omega_\mu^{\ cb}\eeq
In a similar fashion each fermion field can be decomposed into
a real part which is antisymmetric tensor in the frame indices and transforms
as an adjoint under gauge transformations and
an imaginary part which is vector in the local frame and which
gauge transforms
in a similar fashion to the vierbein eg.
\beq
\psi_\mu=\psi^{\ ab}_\mu(x)\sigma^{ab}+i\psi^{\ a}_\mu(x)\gamma^a\eeq
Complexified covariant derivatives acting on such fields also
inherit this same structure.

The interpretation of this Yang-Mills system as
a gravitational theory is further
strengthened by examining the first of the three complex
Yang-Mills Bianchi identities given in
eqn.~\ref{bianchi}. It is straightforward to
show that this leads to the usual two Bianchi identities 
associated with a gravitational theory
\begin{eqnarray}
\epsilon^{\rho\lambda\mu\nu}D_\lambda R_{\mu\nu}&=&0\nonumber\\
\epsilon^{\rho\lambda\mu\nu}[e_\lambda,R_{\mu\nu}]+
\epsilon^{\rho\lambda\mu\nu}D_\lambda T_{\mu\nu}&=&0
\end{eqnarray}
where the covariant
derivatives appearing in the above expressions contain only
the spin connection and we have
introduced the torsion $T_{\mu\nu}=D_{\left[\mu \right.} e_{\left. \nu\right]}$.

Thus the original Yang-Mills model modified to incorporate the
generalized spin connection given in eqn.~\ref{spin} contains
all the ingredients of a gravitational theory. 
The theory contains a spin connection
and a vierbein which satisfy the gravitational Bianchi
identities and an action that is
invariant under both local (Euclidean) Lorentz rotations
and general coordinate transformations
with respect to the background metric $h_{\mu\nu}$. 
Furthermore, the topological nature of
the twisted construction ensures
that the partition function and any
$\cQ$-invariant observables are independent of the choice
of this background metric. The twisted theory is
necessarily supersymmetric which ensures that the theory
possesses additional twisted anti-commuting fields which serve
as superpartners for both the spin connection
and vierbein.

Further evidence supporting this gravitational
interpretation can be 
seen by examining the moduli space of the theory. The first
two vacuum
equations given in eqn.~\ref{moduli} become now\footnote{It is not
clear what is the correct gravitational interpretation of the
3rd equation $D^\mu e_\mu=0$. Its primary role appears to act as
a gauge fixing term to fix the local translational
invariance corresponding to
the imaginary part of the connection. It also may play a role in
selecting out only homogeneous solutions to the vacuum equations} 
\begin{eqnarray}
R_{\mu\nu}^{\ \ ab}-
\frac{2}{l_G^2}\left(e^{\ a}_\mu e^{\ b}_\nu-e^{\ b}_\mu e^{\ a}_\nu\right)&=&0\nonumber\\
T^{\ \ a}_{\mu\nu}=D_\mu e^{\ a}_\nu-D_\nu e^{\ a}_\mu&=&0
\end{eqnarray}
The second relation sets the torsion $T^{\ \ a}_{\mu\nu}$ to zero. 
This together with
the metricity condition $\omega^{\ ab}_\mu=-\omega^{\ ba}_\mu$ is sufficient
to ensure that the $(\omega_\mu,e_\mu)$
system can reconstruct a Riemannian geometry.
Actually we must be careful here; the interpretation of $e^{\ a}_\mu$ as
a vierbein strictly requires that the inverse matrix
$\hat{e}^\mu_{\ a}$ exist with
\beq
\hat{e}^\mu_{\ a} e^{\ b}_\mu=\delta^b_a\eeq
If this is the case we can recast the first of these equations as
\beq
R=\hat{e}^\mu_{\ a}\hat{e}^\nu_{\ b}R^{\ \ ab}_{\mu\nu}=\frac{24}{l_G^2}\eeq
Thus the moduli space corresponds to four dimensional Euclidean
metrics with constant positive
curvature; that is the sphere $S^4$ with radius $l_G$. Actually this
conclusion is strictly valid only for vacuum solutions in which
the scalars $\phi,\phib$ have vanishing expectation values which
may not exhaust all possible classical vacua of the system
\footnote{Note, though, that this model does not possess the
continuum of flat directions found in the usual $U(N)$ theories}. 
Finally, notice that sphere $S^4$ is nothing more than the coset space
$SO(5)/SO(4)$ - that is the space invariant under global
$SO(5)$ transformations modulo local $SO(4)$ rotations.

In general a metric is determined from the vielbeins via the relation
\beq
g_{\mu\nu}=-e^{\ a}_\mu e^{\ a}_\nu=-\Tr(e_\mu e_\nu)\eeq
where the unconventional minus sign reflects our use of the
frame metric $(-1,-1,-1,-1)$.
Thus the moduli space equations imply that this latter
Lorentz invariant operator develops an expectation value in the
classical limit
corresponding to the appearance of an emergent metric.
Notice that while the vacuum energy of the
system is always zero because of supersymmetry the classical limit of the
theory nonetheless leads to a space with non-zero curvature. 

It is likely
that at strong coupling the expectation value of $g_{\mu\nu}$ is zero
and the theory is best described in terms of a strongly
coupled Yang-Mills system of propagating vierbeins and 
spin connections together
with their superpartners as in eqn.~\ref{final}\footnote{Notice that the
the theory as written contains no analog of the Einstein-Hilbert term which is
linear in the curvature. This fact decouples the magnitude of
any effective Newton constant arising from eg supersymmetry
breaking from the scale $l_G$. Such a term could be added by augmenting
the original action by a topological term built from the complex
curvature $\cR_{\mu\nu}$ of the form 
$\int d^4x\;
\epsilon^{\mu\nu\rho\lambda}\epsilon_{abcd}
\cR_{\mu\nu}^{\ \ ab}\cR_{\rho\lambda}^{\ \ cd}$}.

\section{Lorentzian signature}
\label{lorentz}
The generalization to background metrics and tangent spaces with
Lorentz signature is straightforward -- the gamma matrices are taken Lorentzian
and the algebra eqn.~\ref{alg} replaced with 
\begin{eqnarray}
\sigma^{ab}&=&\frac{1}{4}[\gamma^a,\gamma^b]\nonumber\\
\left[\sigma^{ab},\gamma^c\right]&=&\eta^{ca}\gamma^b-\eta^{cb}\gamma^a
\label{algL}\end{eqnarray}
where $\eta^{ab}=\frac{1}{2}\{\gamma^a,\gamma^b\}$ 
corresponds to the Lorentz metric $(+1,-1,-1,-1)$. The algebra of 
the generalized spin connection 
is now isomorphic to the algebra of the group $SO(2,3)$ - 
the isometries of four dimensional anti-de Sitter
space. However, as before, only the Lorentz subgroup $SO(1,3)$
is gauged in the theory. This generalized connection is essentially
the same as that used in \cite{MacDowell:1977jt}. The theory described
here differs from those earlier constructions by virtue of the
twisted supersymmetry which allows us to contract indices using
a background metric while maintaining background independence of
operators which are invariant under the twisted supersymmetry.

To create expressions invariant under this Lorentz symmetry
we need to raise and lower tangent space indices
with the metric $\eta_{ab}$. 
For
example, 
the covariant
derivative of the vierbein and  curvature become
\begin{eqnarray}
D_\mu e_\nu^{\ a}&=&\partial_\mu e_\nu^{\ a}+
\omega_\mu^{\ a}{}_{b} e_\nu^{\ b}\nonumber\\
R_{\mu\nu\ b}^{\ \ a}&=&\partial_\mu\omega_\nu^{\ a}{}_{b}-
\partial_\nu\omega_\mu^{\ a}{}_{b}+
\omega_\mu^{\ a}{}_{c}\omega_\nu^{\ c}{}_{b}-
\omega_\nu^{\ a}{}_{c}\omega_\mu^{\ c}{}_{b}
\end{eqnarray}
while the metric is now given by
\beq
g_{\mu\nu}=e^{\ a}_\mu e_{\ a}{}_\nu=\eta_{ab}e^{\ a}_\mu e^{\ b}_\nu\eeq
The action can be rendered locally Lorentz invariant by ensuring that
all tangent space indices are contracted using $\eta_{ab}$ in the
usual way. 
Again, the torsion vanishes on
the moduli space and the spin connection and vierbein
combine to yield a constant curvature
space. 
In this case the metric describes the coset space
$SO(2,3)/SO(1,3)$ or four dimensional anti-de Sitter space with
curvature $\frac{24}{l_G^2}$.

\section{Five dimensional AdS}
\label{AdS5}
As was shown in \cite{Catterall:2007kn} the Marcus twist of ${\cal N}=4$
Yang-Mills is most succinctly written as the dimensional reduction of
a five dimensional gauge theory comprising
a complex gauge field $\cA_\mu, \mu=1 \ldots 5$ and a multiplet
of twisted fermions $(\eta,\psi_\mu,\chi_{\mu\nu})$.
The five dimensional action takes the form
\beq
S=\frac{1}{g_5^2}\left(\cQ\int d^5x\sqrt{h}\Lambda-
\frac{1}{4}\int d^5x\;\Tr\epsilon^{\mu\nu\rho\lambda\delta}
\chi_{\mu\nu}\cDb_\rho\chi_{\lambda\delta}\right)\label{5action}\eeq
where the $\cQ$-exact term is given by
\beq
\Lambda=\int
\Tr\left(\chi^{\mu\nu}\cF_{\mu\nu}+\eta [ \cDb^\mu,\cD_\mu ]-\frac{1}{2}\eta
d\right)
\eeq
where $\cF_{\mu\nu}$ is the complex Yang-Mills curvature associated
to the five dimensional complex connection $\cA_\mu$ 
and we have introduced now a five dimensional 
background metric $h_{\mu\nu}(x)$. 
As before
the nilpotent supersymmetry acts on the fields as follows
\begin{eqnarray}
\cQ\; \cA_\mu&=&\psi_\mu\nonumber\\
\cQ\; \psi_\mu&=&0\nonumber\\
\cQ\; \cAb_\mu&=&0\nonumber\\
\cQ\; \chi_{\mu\nu}&=&-\cFb_{\mu\nu}\nonumber\\
\cQ\; \eta&=&d\nonumber\\
\cQ\; d&=&0
\end{eqnarray}
The second $\cQ$-closed term is supersymmetric on account of the
generalized Bianchi identity 
\beq
\epsilon^{\mu\nu\rho\lambda\delta}\cD_\rho \cF_{\lambda\delta}=0
\label{5bianchi}
\eeq
The definitions of the complexified covariant derivatives follow those
of the four dimensional theory. 

The transition to a theory of gravity is performed as before; we expand
the connections not on the Lie algebra of $U(N)$ but that of
the {\it five dimensional} anti-de Sitter algebra $SO(2,4)$\footnote{As for
four dimensions we may also use the algebra $so(6)$ which will
lead to a $S^5$ vacuum state}. Again this algebra separates into
a piece corresponding to the five dimensional Lorentz group 
$SO(1,4)$ which is
gauged and a translational component representing the 5-bein.
Again, the complexified Bianchi identities yield the two
Bianchi identities of a gravitational theory. And once again the expectation
values of topological observables are independent of smooth
deformations of the background metric $h_{\mu\nu}$.

The classical vacua are now given simply by the conditions
\begin{eqnarray}
R_{\mu\nu}^{\ \ ab}-\frac{2}{l_G^2}\left(
e^{\ a}_\mu e^{\ b}_\nu-e^{\ b}_\mu e^{\ a}_\nu\right)&=&0\nonumber\\
T^{\ \ a}_{\mu\nu}=D_\mu e^{\ a}_\nu-D_\nu e^{\ a}_\mu&=&0\nonumber\\
D^\mu e_\mu^{\ a}&=&0
\label{5dmod}
\end{eqnarray}
where $R_{\mu\nu}$ is the curvature associated to the spin connection
$\omega_\mu$ and $e_\mu$ is the 5-bein. As for four dimensions
these equations
can be interpreted as yielding five dimensional anti-de Sitter space
with radius $l_G$.

The fermions appearing in this
construction bear some resemblance to 
those of five dimensional $\cN=4$ supergravity \cite{Tanii:1998px}.
For example, the theory contains a fermionic superpartner of
the generalized spin connection  $\psi_\mu$
whose component fields $\psi_\mu^{\ ab},\psi_\mu^{\ a}$
can be thought of as resulting from the four degenerate
gravitinos expected in $\cN=4$ supergravity after
twisting the frame indices with the associated flavor
symmetry. Similarly, the fermions $\eta^a,\eta^{ab}$ can be mapped into
the four spinors of such a theory after the same
twisting of tangent space indices with
$\cN=4$ flavor indices. It is also interesting
to observe that the {\it global} background $\cQ$ supersymmetry
transforms the 5-bein into the twisted gravitino as required
by {\it local} supersymmetry \cite{vanNieuwenhuizen:2004rh}.

\section{Lattice construction}

The five dimensional construction described
above is the basis for
recent efforts to derive lattice actions for ${\cal N}=4$ super Yang-Mills
theory 
which retain an exact supersymmetry at non zero lattice spacing.
For completeness we summarize this construction here
specializing
where necessary to the gravitational case. For more details the
reader is referred to \cite{Catterall:2007kn,Kaplan:2005ta,Damgaard:2008pa}.

Clearly to make contact with a twist of ${\cal N}=4$ in four dimensions
we must dimensionally reduce this theory along the 5th direction.
This will yield the complex scalar $\phi=A_5+iB_5$ and its 
superpartner $\etab$ that appeared in the continuum construction
given in section~\ref{gauge}.
Similarly
the 10 five dimensional fermions $\chi_{\mu\nu},\mu=1\ldots 5$ 
naturally decompose into the 6 fields of
a 2-form in four dimensions and the vector $\psib_\mu,\mu=1\ldots 4$.  
However the lattice action is most simply
expressed in language of the five dimensional theory and we will
hence use that notation in what follows - thus all indices will
be taken as varying in the
range
$\mu,\nu=1\ldots 5$.

The transition to the lattice from the continuum theory 
requires a number of steps. The first, and most important,
is to replace the continuum generalized spin connection $\cA_\mu(x)$
which now takes its value in the four dimensional anti-de Sitter algebra
by a set of appropriate complexified Wilson
link fields which are the exponentials of these fields
$\cU_\mu(x)=e^{\cA_\mu(x)},\mu=1\ldots 4$. These lattice fields 
are taken to be
associated with unit length vectors in the coordinate
directions $\bmu$ in an abstract
four dimensional hypercubic lattice. In contrast the continuum
scalar field $\cA_5$ takes its values only in the algebra and is placed on
the body diagonal
$\bmu_5=(-1,-1,-1,-1)$ of the unit hypercube\footnote{A more symmetrical
treatment of the scalar field is also possible and leads
to an $A_4^*$ lattice - see \cite{Catterall:2007kn,
Kaplan:2005ta}}.  
Notice that the sum of these five
basis vectors is zero. This is very
important for gauge invariance of the lattice model. By supersymmetry the fermion
fields $\psi_\mu,\mu=1\ldots 5$ lie on the same link as their
bosonic superpartners. In contrast the scalar
fermion $\eta$ is associated with the sites of the lattice and the
tensor fermions $\chi_{\mu\nu},\mu=1\ldots 5$ 
with the link running from $\bmu+\bnu$ down to
the origin of the hypercube (remember that
the set of vectors $\bmu$,$\bnu$ runs over
the unit vectors in the coordinate directions together with the body diagonal
$\bmu_5$). These are the same link assignments that occur in the
lattice constructions of sixteen
supercharge $U(N)$ theories 
\cite{Catterall:2007kn,Kaplan:2005ta,Damgaard:2008pa}.

The construction then posits that
all link fields transform as bifundamental
fields under gauge transformations
\begin{eqnarray}
\eta(\bx)&\to& G(\bx)\eta(\bx) G^\dagger(\bx)\nonumber\\
\psi_\mu(\bx)&\to& G(\bx)\psi_\mu(\bx) G(\bx+\bmu)\nonumber\\
\chi_{\mu\nu}(\bx)&\to&G(\bx+\bmu+\bnu)\chi_{\mu\nu}(\bx)G^\dagger(\bx)\nonumber\\
\cU_\mu(\bx)&\to&G(\bx)\cU_\mu(\bx)G^\dagger(\bx+\bmu)\nonumber\\
\cUb_\mu(\bx)&\to&G(\bx+\bmu)\cUb_\mu(\bx)G^\dagger(\bx)
\label{gaugetrans}
\end{eqnarray}
The supersymmetric and gauge invariant
lattice action which corresponds to
eqn.~\ref{action} then takes a very similar form
to its continuum counterpart in eqn.~\ref{5action}
\beq
S=\frac{1}{g^2}\left(\cQ\sum_x \Lambda-
\frac{1}{4}\sum_x\;\Tr \epsilon^{\mu\nu\rho\lambda\delta}
\chi_{\mu\nu}\cDb^{(-)}_\rho\chi_{\lambda\delta}\right)\label{laction}\eeq
where the $\cQ$-exact term is given by
\beq
\Lambda=
\Tr\left(\chi_{\mu\nu}\cR_{\mu\nu}+\eta \cDb^{(-)}_\mu\cU_\mu-\frac{1}{2}\eta
d\right)
\eeq
and the form of the supersymmetry transformations are the same as the
continuum theory
\begin{eqnarray}
\cQ\; \cU_\mu&=&\psi_\mu\nonumber\\
\cQ\; \psi_\mu&=&0\nonumber\\
\cQ\; \cUb_\mu&=&0\nonumber\\
\cQ\; \chi_{\mu\nu}&=&\cR_{\mu\nu}^\dagger\nonumber\\
\cQ\; \eta&=&d\nonumber\\
\cQ\; d&=&0
\end{eqnarray}
The lattice field strength is given by
\beq
\cR_{\mu\nu}=\cD^{(+)}_\mu \cU_\nu(\bx)=
\cU_\mu(\bx)\cU_\nu(\bx+\bmu)-\cU_\nu(\bx)\cU_\mu(\bx+\bnu)\eeq
This lattice action is invariant under local $SO(1,3)$ gauge
transformations. It is also free of fermion doubling problems -- see the
discussion in \cite{Catterall:2007kn}.
The covariant lattice difference operators appearing in these expressions
are defined by
\begin{eqnarray}
\cD^{(+)}_\mu f_\nu(\bx)&=&\cU_\mu(\bx)f_\nu(\bx+\bmu)-f_\nu(\bx)U_\mu(\bx+\bmu)\\
\cD^{(-)}_\mu f_\mu(\bx)&=&f_\mu(\bx)\cUb_\mu(\bx)-\cUb_\mu(x-\bmu)f_\mu(\bx-\bmu)
\end{eqnarray}  
Notice that $\cD^{(+)}$ acts like an exterior derivative and promotes a
lattice $p$-form to a $(p+1)$-form as evidenced by its gauge
transformation property. Similarly $\cD^{(-)}$ maps a lattice
$p$-form to a $(p-1)$-form corresponding to the adjoint of the
exterior derivative.
Finally, the derivative appearing in the $\cQ$-closed term is given by
\beq
\cDb^{(-)}_\rho\chi_{\mu\nu}(\bx)=\chi_{\mu\nu}(\bx)\cUb_\rho(\bx-\brho)-
\cUb_\rho(\bx+\bmu+\bnu-\brho)\chi_{\mu\nu}(\bx-\brho)
\eeq
It is straightforward to show that with this
definition the complex Yang-Mills Bianchi identity is satisfied
{\it exactly} in the lattice theory \cite{Catterall:2007kn}.

Notice that this lattice theory bears some similarity to
spin foam formulations of loop quantum gravity
\cite{Smolin:2004sx}. What is different is that the
underlying lattice structure is fixed corresponding to
an implicit choice of a flat background metric and the Wilson loop
variables take values in the anti-de Sitter group with only the
Lorentz subgroup being gauged.

\section{Discussion}

In this paper we have constructed a topological gravity theory of
Yang-Mills type by modifying the Marcus twist
of $\cN=4$. The key to the construction is to embed the spin
connection and vierbein into a complexified
connection whose generators taken together span
the anti-de Sitter algebra. However only the Lorentz
subalgebra is gauged. The complexified Yang-Mills Bianchi identities
then yield the usual two Bianchi identities of the gravitational
theory and the classical vacua are shown to correspond to
anti-de Sitter space. It is only in this
classical limit that a metric tensor can be clearly identified.

Away from the classical limit
the torsion
is non-zero, the spin connection and vierbein are 
independent, interacting 
fields and the metric tensor constructed from
the vierbeins most probably has a
vanishing expectation value. 

The topological construction
ensures that the partition function of the model is
independent of the background metric and indeed may be computed
exactly in the semi-classical approximation.
Furthermore, 
the twisted supersymmetry guarantees that the vacuum energy
is zero while
the curvature of the space, being a parameter of
the moduli space, should suffer no quantum corrections. 

The
appearance in this model of a moduli space corresponding
to flat connections is similar to
previous constructions of two dimensional topological
supergravity \cite{Montano:1990ru}. It is also likely that
the theory described here is related to a supersymmetric
extension of BF-theory such as that proposed in 
\cite{Smolin:2003qu,Freidel:2005ak,KowalskiGlikman:2008fj}. In these
constructions the topological terms are supplemented by
additional cubic interactions which ensure the theory reduces
to General Relativity in an appropriate limit.  Further evidence in
favor of this comes from the work of
Blau and Thompson who showed that the twisted $\cN=4$ theory
considered here with
$U(N)$ gauge group could be obtained as a deformation of
super BF theory \cite{Blau:1996bx}.

While $\cQ$-invariant observables in this
continuum theory are background independent this
will not be true of general gauge invariant operators. 
This appears to be a barrier preventing a non-topological
gravity interpretation of the theory. One way to
proceed is to simply give up on the requirement
of general background independence and consider the Yang-Mills theory as
defined on a flat base space. The gravitational interpretation of
the theory then rests upon the observation that the bosonic fields of
the model
satisfy the gravitational Bianchi identities and that the 
vacua can be interpreted as a classical spacetime. 
It would 
be very interesting to see whether small fluctuations of the
coupled vierbein/spin connection system around this
vacuum state could be interpreted in terms of fluctuations of
an emergent metric tensor and, in particular, how to reconcile
such a picture with the Weinberg-Witten theorem \cite{Weinberg:1980kq}.
Notice that the latter theorem, which prohibits the appearance of
a composite massless graviton, requires the existence of a 
Poincar\'{e}
covariant stress energy tensor. In the theory described here
any gravitational fluctuations
occur relative to an AdS spacetime and may yield an emergent
graviton with an effective mass determined by the scale $l_G$ thereby
evading the theorem.

In the case of a flat background we have also written
down a lattice theory which, in the
naive continuum limit, approximates this
continuum theory arbitrarily well. 
This lattice theory is both gauge invariant and preserves the
scalar supersymmetry and it
can be thought of as yielding a non-perturbative 
formulation of the theory suitable
for numerical simulations \cite{Catterall:2008dv}. 

\acknowledgments The author is supported in part by DOE grant
DE-FG02-85ER40237. He gratefully acknowledges Poul Damgaard, Joel Giedt,
David B. Kaplan, 
Don Marolf,  
So Matsuura, Lee Smolin, Mithat \"{U}nsal and Toby Wiseman for
a critical reading of the manuscript.

%
\bibliographystyle{JHEP}
\bibliography{AdS2}

\providecommand{\href}[2]{#2}\begingroup\raggedright\begin{thebibliography}{10}

\bibitem{Donoghue:1994dn}
J.~F. Donoghue, {\it {General relativity as an effective field theory: The
  leading quantum corrections}},  {\em Phys. Rev.} {\bf D50} (1994) 3874--3888,
  [\href{http://xxx.lanl.gov/abs/gr-qc/9405057}{{\tt gr-qc/9405057}}].

\bibitem{vanNieuwenhuizen:2004rh}
P.~van Nieuwenhuizen, {\it {Supergravity as a Yang-Mills theory}},
  \href{http://xxx.lanl.gov/abs/hep-th/0408137}{{\tt hep-th/0408137}}.

\bibitem{Anselmi:1992tj}
D.~Anselmi and P.~Fre, {\it {Twisted N=2 supergravity as topological gravity in
  four- dimensions}},  {\em Nucl. Phys.} {\bf B392} (1993) 401--427,
  [\href{http://xxx.lanl.gov/abs/hep-th/9208029}{{\tt hep-th/9208029}}].

\bibitem{Marcus}
N.~Marcus, {\it The other topological twisting of n=4 yang-mills},  {\em Nucl.
  Phys.} {\bf B452} (1995) 331--345,
  [\href{http://xxx.lanl.gov/abs/hep-th/9506002}{{\tt hep-th/9506002}}].

\bibitem{Kapustin:2006pk}
A.~Kapustin and E.~Witten, {\it {Electric-magnetic duality and the geometric
  Langlands program}},  \href{http://xxx.lanl.gov/abs/hep-th/0604151}{{\tt
  hep-th/0604151}}.

\bibitem{Unsal:2006qp}
M.~Unsal, {\it Twisted supersymmetric gauge theories and orbifold lattices},
  {\em JHEP} {\bf 10} (2006) 089,
  [\href{http://xxx.lanl.gov/abs/hep-th/0603046}{{\tt hep-th/0603046}}].

\bibitem{Catterall:2007kn}
S.~Catterall, {\it {From Twisted Supersymmetry to Orbifold Lattices}},  {\em
  JHEP} {\bf 01} (2008) 048, [\href{http://xxx.lanl.gov/abs/0712.2532}{{\tt
  0712.2532}}].

\bibitem{Ashtekar:1986yd}
A.~Ashtekar, {\it {New Variables for Classical and Quantum Gravity}},  {\em
  Phys. Rev. Lett.} {\bf 57} (1986) 2244--2247.

\bibitem{MacDowell:1977jt}
S.~W. MacDowell and F.~Mansouri, {\it {Unified Geometric Theory of Gravity and
  Supergravity}},  {\em Phys. Rev. Lett.} {\bf 38} (1977) 739.

\bibitem{Tanii:1998px}
Y.~Tanii, {\it {Introduction to supergravities in diverse dimensions}},
  \href{http://xxx.lanl.gov/abs/hep-th/9802138}{{\tt hep-th/9802138}}.

\bibitem{Kaplan:2005ta}
D.~B. Kaplan and M.~Unsal, {\it A euclidean lattice construction of
  supersymmetric yang- mills theories with sixteen supercharges},  {\em JHEP}
  {\bf 09} (2005) 042, [\href{http://xxx.lanl.gov/abs/hep-lat/0503039}{{\tt
  hep-lat/0503039}}].

\bibitem{Damgaard:2008pa}
P.~H. Damgaard and S.~Matsuura, {\it {Geometry of Orbifolded Supersymmetric
  Lattice Gauge Theories}},  {\em Phys. Lett.} {\bf B661} (2008) 52--56,
  [\href{http://xxx.lanl.gov/abs/0801.2936}{{\tt 0801.2936}}].

\bibitem{Smolin:2004sx}
L.~Smolin, {\it {An invitation to loop quantum gravity}},
  \href{http://xxx.lanl.gov/abs/hep-th/0408048}{{\tt hep-th/0408048}}.

\bibitem{Montano:1990ru}
D.~Montano, K.~Aoki, and J.~Sonnenschein, {\it {TOPOLOGICAL SUPERGRAVITY IN
  TWO-DIMENSIONS}},  {\em Phys. Lett.} {\bf B247} (1990) 64--70.

\bibitem{Smolin:2003qu}
L.~Smolin and A.~Starodubtsev, {\it {General relativity with a topological
  phase: An action principle}},
  \href{http://xxx.lanl.gov/abs/hep-th/0311163}{{\tt hep-th/0311163}}.

\bibitem{Freidel:2005ak}
L.~Freidel and A.~Starodubtsev, {\it {Quantum gravity in terms of topological
  observables}},  \href{http://xxx.lanl.gov/abs/hep-th/0501191}{{\tt
  hep-th/0501191}}.

\bibitem{KowalskiGlikman:2008fj}
J.~Kowalski-Glikman and A.~Starodubtsev, {\it {Effective particle kinematics
  from Quantum Gravity}},  {\em Phys. Rev.} {\bf D78} (2008) 084039,
  [\href{http://xxx.lanl.gov/abs/0808.2613}{{\tt 0808.2613}}].

\bibitem{Blau:1996bx}
M.~Blau and G.~Thompson, {\it {Aspects of $N_{T}\geq 2$ Topological Gauge
  Theories and D- Branes}},  {\em Nucl. Phys.} {\bf B492} (1997) 545--590,
  [\href{http://xxx.lanl.gov/abs/hep-th/9612143}{{\tt hep-th/9612143}}].

\bibitem{Weinberg:1980kq}
S.~Weinberg and E.~Witten, {\it {Limits on Massless Particles}},  {\em Phys.
  Lett.} {\bf B96} (1980) 59.

\bibitem{Catterall:2008dv}
S.~Catterall, {\it {First results from simulations of supersymmetric
  lattices}},  \href{http://xxx.lanl.gov/abs/0811.1203}{{\tt 0811.1203}}.

\end{thebibliography}\endgroup
%

\end{document}